1 **Pacing Early Mars fluvial activity at Aeolis Dorsa: Implications for Mars**

2 **Science Laboratory observations at Gale Crater and Aeolis Mons**

4 Edwin S. Kite[a] (ekite@caltech.edu), Antoine Lucas[a], Caleb I. Fassett[b]

5 [a] Caltech, Division of Geological and Planetary Sciences, Pasadena, CA 91125

6 [b] Mount Holyoke College, Department of Astronomy, South Hadley, MA 01075

8 **Abstract:** The impactor flux early in Mars history was much higher than today, so sedimentary

9 sequences include many buried craters. In combination with models for the impactor flux,

10 observations of the number of buried craters can constrain sedimentation rates. Using the

11 frequency of crater-river interactions, we find net sedimentation rate ≲20-300 μm/yr at Aeolis

12 Dorsa. This sets a lower bound of 1-15 Myr on the total interval spanned by fluvial activity

13 around the Noachian-Hesperian transition. We predict that Gale Crater's mound (Aeolis Mons)

14 took at least 10-100 Myr to accumulate, which is testable by the Mars Science Laboratory.

16 **1. Introduction.**

17 On Mars, many craters are embedded within sedimentary sequences, leading to the

18 recognition that the planet's geological history is recorded in "cratered volumes", rather than

19 just cratered surfaces (Edgett and Malin, 2002). For a given impact flux, the density of craters

20 interbedded within a geologic unit is inversely proportional to the deposition rate of that

21 geologic unit (Smith et al. 2008). To use embedded-crater statistics to constrain deposition

22 rate, it is necessary to distinguish the population of interbedded craters from a (usually much

23 more numerous) population of craters formed during and after exhumation. However, on

24 Mars, erosion can exhume intact impact craters complete with ejecta blankets from beneath

25 hundreds of meters of overlying sediment (e.g., Edgett, 2005). Variations in burial, exhumation,

viscous relaxation, and erosion produce varying crater-preservation styles (Figure 1). Because of the difficulty of determining which craters are syndepositional, we do not know of any previous measurements of fluvial sedimentation rates on Mars using the frequency of embedded impact craters.

Geological evidence suggests a period of enhanced precipitation-fed fluvial erosion created regionally-integrated highland valley networks around the Noachian-Hesperian boundary ~3.6-3.7 Ga (Irwin et al., 2005). Regionally-integrated fluvial activity ceased planetwide at roughly the same time, perhaps even synchronously (Fassett et al., 2011), although limited and/or localized fluvial activity continued afterward (e.g., Grant & Wilson, 2011; Mangold et al., 2012). Lake-basin hydrology disfavors valley-network formation by a single deluge (Barnhart et al., 2009; Matsubara et al., 2011). Improved constraints on the pace and persistence of fluvial activity during the Noachian-Hesperian transition are required to understand early Mars climate and habitability. It remains unclear whether the fluvial geomorphology of the Noachian-Hesperian transition is a palimpsest of transient events triggered by volcanic eruptions or impacts (Segura et al., 2008; Kite et al., 2011; Wordsworth et al., 2012), or alternatively records a sustained (>>1 Kyr) wet interval(s) caused by unusual orbital conditions (Kite et al., 2012a), an early greenhouse (Sagan & Mullen, 1972), or an impact-triggered excursion to a warm stable state (Segura et al., 2012).

Here we use interactions between craters and fluvial deposits to determine sedimentation rate within a Late Noachian/Early Hesperian sedimentary basin containing numerous fluvial deposits. A river flowing over a heavily cratered sedimentary landscape will be frequently diverted by crater rims and/or deposit sediment in pools corresponding to the crater interiors (Figure 1). These craters, and craters partly overlain by river deposits, are readily identified as being syndepositional.

**2. River-crater interactions.**

We search for river-crater interactions within exceptionally numerous and well-preserved fluvial channel deposits exhumed by erosion in the Aeolis Dorsa region, formerly termed Aeolis-Zephyria Planum (Burr et al., 2009, 2010; note that we use Aeolis Dorsa to refer to the formally defined region, not just the ridges within that region). The area of our search (~6S, 152E; Figure 2) is stratigraphically ~500m below a surface with an Early Hesperian, minimum model age of 3.69(+0.05/-0.07) Ga (Zimbelman & Scheidt, 2012, using the Ivanov et al., 2001 production function and the Hartmann & Neukum, 2001 chronology function). It forms part of an eroded deposit that is thought to be Late Noachian or Early Hesperian (Irwin et al., 2004). These dates correlate the Aeolis rivers to the Noachian-Hesperian transition and to the lower Gale Crater mound (Fassett & Head, 2008; Kerber & Head, 2010; Thomson et al, 2011; Zimbelman & Scheidt, 2012); they allow for the rivers to either predate, or correlate with, the lower Gale Crater mound.

Packing density of channel deposits within the stratigraphy is comparable to Earth fluviodeltaic basins (e.g., Gouw, 2008). The Martian rivers are qualitatively similar to meandering-river deposits on Earth. Interactions with craters are, therefore, easily recognized as anomalous. Impact craters are identified by random distribution across the landscape, obliteration of older geological structures within the transient cavity, and especially circularity and upturned rims. Interfluve material would be interpreted as the fine-grained deposits of rare floods if this were a basin on Earth. However, in the absence of grain-size data or diagnostic levee-breach features, wind-blown dust and silt cannot be ruled out (Haberlah et al., 2010). From local relationships the stratigraphic thickness of deposits containing fluvial channels, $\Delta z$, is > 100m. Channel deposits are exposed over a ~1km elevation range, but postdepositional modification complicates reconstruction of $\Delta z$ (Lefort et al., 2012). A likely

75   lower bound on $\Delta z$ is the difference between the modern surface and a surface interpolated
76   inward from low points surrounding the fluvial region. This gives $\Delta z \gtrsim 300$m.

77   17 exhumed craters are found (Supplementary Table) at multiple stratigraphic levels
78   within 2100 km² of fluvial deposits surveyed with CTX images (5-6 m/pixel). Identifications
79   were checked where possible with higher-resolution images (HiRISE or MOC). In addition to
80   these features that are definitely identified as embedded within the stratigraphy and having
81   definite impact crater morphology, an additional 43 candidates were found. Crater diameters
82   were obtained by fitting a circle to the visible arc of the crater edge. Embedded-crater size-
83   frequency distributions for $D$ >250m ($D$ is diameter) have a cumulative power-law slope
84   slightly shallower than –2, ~1 less than the production-function slope of –3 in this range. This
85   is expected for a crater population embedded within a volume (Yielding et al., 1996); a pristine
86   crater population on a geologically stable surface would parallel the production function.
87   Craters $D \lesssim 250$ m are still further underrepresented, which we interpret as the result of
88   survey incompleteness or poor exhumed-crater preservation at small sizes.

89

90   **3. Constraints on sedimentation rate and fluvial timescales.**

91   We make the following initial assumptions –

92      (i) Cratering is a Poisson process with an initial crater depth, $d \approx 0.2D$ for $D$ < 1km (Melosh,
93      1989).

94      (ii) Erosion does not preferentially expose craters.

95      (iii) During deflation of the deposit an embedded crater is invisible until the deflation
96      surface reaches the embedded-crater rim. It is then visible at its original diameter until the
97      deflation surface reaches the level of the bottom of the crater.

98   We discuss possible violations of these assumptions later.

99    The flux of impact craters at the time the deposit was forming, $f$ ($D > D_i$), is obtained

100   using the crater-production and crater-chronology functions recommended by Werner &

101   Tanaka (2011). Differences between the Hartmann and Ivanov/Neukum-Hartmann functions

102   lead to <20% disagreement in resurfacing rate for the size-range of craters used here (Figure

103   3), which is unimportant compared to other uncertainties. The expected number of embedded

104   craters, $N_{cr}$, is given by

105                        $N_{cr}$ ($D > D_i$) = $f$ ($D > D_i$) ($d/D$) $D$ $a$ / $S$

106   where $a$ is count area, and $S$ is accumulation rate. Excluding $D < 290$m craters and assuming an

107   age in the range 3.7 – 3.9 Ga, least-squares fitting of accumulation rates to the data (Figure 3)

108   gives $S$ = 20-80 μm/yr (or 70-300 μm/yr including all candidates). Therefore, $S \approx$ 20-300

109   μm/yr for the range of likely ages.

110   The most important uncertainty is $f$. The Aeolis Dorsa river deposits were emplaced during

111   a period of higher crater flux $f$ when rapid changes in $f$ are also possible, potentially associated

112   with the Late Heavy Bombardment. Therefore, small changes in the age of the deposit may lead

113   to large changes in $S$ (Werner & Tanaka, 2011). Crater-chronology functions are defined based

114   on the lunar sample collection, whose interpretation is somewhat model-dependent, and the

115   translation of these data to Mars is challenging (e.g., Ivanov, 2001).

116   Other uncertainties tend to lead to an underestimate of exhumed-crater frequency. (a) If

117   channel belts aggraded faster than interfluves and are erosionally resistant, and erosion is by

118   vertical downcutting uncorrelated with laterally adjacent terrain, then at any given time the

119   modern deflation surface will preferentially expose the erosion-resistant, crater-deficient

120   units. In this case our procedure would underestimate the frequency of craters per unit volume

121   averaged over the basin. On Earth, channel belts always aggrade faster than their floodplains

122   on interannual timescales, but levee breaches and avulsions maintain constant aggradation

123   rate across the floodplain averaged over depths greater than ~1 channel depth (Mohrig et al.,

2000). Channel-width measurements (Burr et al., 2010), standard fluvial scaling relations, and HiRISE DTM measurements of negative-relief channel depths indicate that Aeolis channel depths should be small compared to the original depth of the craters in our count (i.e. channel depth << 30-60m). Therefore lateral gradients in sedimentation rate are unlikely to be important if the interfluves are floodplain deposits. (b) Cliff-forming units that are very resistant to vertical abrasion will be removed by lateral mass wasting as surrounding weaker material is eroded. In this case, no embedded craters within the cliff-forming unit will be included in the count: craters on top of the cliff-forming unit will be hard to distinguish from relatively recent synerosional craters, and craters within the unit will be blanketed by talus throughout the erosion process. The effective count area then scales with the perimeter of the cliff-forming unit, rather than with its area. (c) $S$ would halve in the extreme case that all craters initially have a secondary-like $d/D$ (i.e., ~0.1). $d/D$ is currently <0.1 in our HiRISE DTMs, but this could be due to incomplete erosion or recent infilling. (d) Supposing small craters were all erosionally resistant and formed mesas as tall as their diameter, asumption (iii) would be violated and a surface count would overestimate the true crater density. However, our CTX and HiRISE DTMs show that small exhumed craters are not locally highstanding in this region, although they do tend to be preserved with rims intact, so (iii) is probably a good approximation. (e) We assume craters are either present at full diameter or eliminated completely. (f) Finally, we neglect possible erosion of craters *during* the period of net accumulation.

Because these errors tend to lead to an undercount of the number of craters that formed during the interval of fluvial deposition, we interpret our data as a lower limit on time for accumulation. Assuming $\Delta z \gtrsim 300$m, the range of minimum deposition timescale is $\Delta z/S$ ~1-15 Ma for the range of likely ages (or 0.5-30 Ma assuming a wider range of age uncertainty, from 3.6-4.0 Ga).

## 4. Discussion.

The simplest interpretation of these data is fluvial aggradation at rates comparable to Earth (fluvial aggradation rates ranging from 50-600 μm/yr are compiled by Miall, 2012). Aeolis Dorsa sedimentation cannot be distinguished from later sedimentation on the basis of sedimentation rate alone. Putzig et al. (2009) correlate radar reflectors within the North Polar Layered Deposits to 0-4 Mya obliquity cycles and obtain $S$ ~1 mm water ice/yr. Similarly, Lewis et al. (2008) correlate bed:bundle ratios in Becquerel Crater at 22°N to Milankovitch beats and obtain $S$ ~30 μm/yr. Typical equatorial mound rhythmic-layer thicknesses of 3-20m (Lewis et al., 2010) imply accumulation at 20-200 μm/yr if forced by obliquity cycles (0.12 Myr), or 100-800 μm/yr if forced by precession (0.025 Myr effective period at the equator). Rhythmites have been hypothesized to be relatively young (Grotzinger & Milliken, in press). Modern gross sedimentation rate on Mars is 10-100 μm/year from dust storms (Drube et al., 2010).

On Earth, mean sedimentation rate frequently decreases with increasing measurement duration (Jerolmack & Sadler, 2007) as a result of power-law fluctuations of the boundary between erosion and deposition (Schumer & Jerolmack, 2009). This "Sadler effect" is ubiquitous at short timescales near coasts but is less relevant on the longest timescales, or where erosion is unimportant (Jerolmack & Sadler, 2007). It is not clear whether the Sadler effect should apply to Mars sediments. Martian weather is remarkably predictable on both synoptic and interannual timescales, probably because of the lack of a large energy capacitor analogous to Earth's ocean (Read & Lewis, 2004). Therefore, we might speculate that Mars' sedimentary record, as the imprint of the atmosphere on rocks, is less chaotic than its Earth counterpart. Consistent with this, quasi-periodic bedding is common on Mars (Lewis et al., 2010) and angular unconformities are rare. Our data hint that exhumed craters are concentrated at a few stratigraphic levels, consistent with omission surfaces.

174   The sediment source for the Aeolis Dorsa deposits is uncertain. It is possible that sediment

175   was fluvially transported from highlands to the south. However, the volume of valley networks

176   draining toward Aeolis Dorsa appears to be much smaller than the volume of the clastic wedge,

177   and no complete transport pathways are visible. It is conceivable that the Aeolis Dorsa deposits

178   are fluvially reworked ancient highlands crust, which would require that the dichotomy

179   boundary was once further to the north. But crater-floor tilts suggest that the dichotomy

180   boundary was in place near its current location prior to the late Noachian (Watters et al.,

181   2007). An attractive possibility is that the fluvial deposits are reworked from relatively weak

182   aeolian or niveoaeolian deposits accumulating at the highland-lowland boundary (Irwin,

183   2004).

184   The relatively low embedded-crater frequency is in line with low Platinum Group Element

185   concentrations in 3.8 Ga metasediments from Earth (Anbar et al., 2001).

186

187   **5. Implications for ancient climate and MSL's mission to Gale Crater.**

188   A lower bound of 1-15 Myr for the total interval spanned by fluvial deposition rules out basin-

189   filling by a single catastrophic episode, and is consistent with hydrologic and total-erosion

190   estimates for fluvial activity around the Noachian-Hesperian boundary (Barnhart et al., 2009;

191   Hoke et al., 2011). Possible climate regimes include multiple climate transients (Segura et al.

192   2008, Wordsworth et al. 2012), or intermittent precipitation-fed runoff over $\geq 10^5$ yr (Barnhart

193   et al., 2009). The implied total time interval recorded by Gale-Aeolis sediments is $\geq 10^{7-8}$ yr.

194   The sedimentary rock record of Mars appears to record a small fraction of Mars history,

195   perhaps because surface liquid water was necessary for lithification and was only available

196   intermittently (the wet-pass filter hypothesis; Moore, 1990, Knoll et al., 2008, Andrews-Hanna

197   & Lewis, 2011, Kite et al., 2012a).

Based on observed and candidate embedded-crater frequency in Aeolis Dorsa, we predict a mean Aeolis Mons sedimentation rate of 20-300 µm/yr including nondepositional intervals. This is testable with MSL measurements of cosmogenic noble gases (e.g., Shuster et al., 2012), meteoritic Ni (Yen et al., 2006), meteoritic organic matter (OM), and small embedded craters. Mars Hand Lens Imager's (MAHLI's) 14µm resolution permits identification of ≥50µm-thick varves, if they exist.

Sediment accumulation rate affects the rate at which OM from meteoritic infall is introduced to the record – the "meteoritic background level" (Summons et al., 2011). Preservation of OM introduced at the surface is affected by time-to-burial to a depth of order 1m. During this time, OM is vulnerable to degradation by radiation, atmospheric oxidants, and/or UV, to an extent that depends on the unknown redox state, composition and thickness of Early Mars' atmosphere (Pavlov et al., 2012). On Earth, overbank environments and oxbow lakes in meander belts are favored among subaerial environments for preserving organic matter (Summons et al., 2011).

River deposits exist at the Gale mound, but those clearly identifiable from orbit appear to postdate the accumulation of the lower unit of Aeolis Mons. It has been argued that the primary sediment source for the Gale mound is airfall (Pelkey et al., 2004; Thomson et al., 2011; Kite et al., 2012b). If this is true, then the extrapolation of the Aeolis Dorsa sedimentation rate to Aeolis Mons depends on the assumption that river sediment was supplied by aeolian processes (e.g., Irwin et al., 2004; Haberlah, 2010).

Future work might use MSL data to assess the relation between exposure time, crater frequency, OM preservation, and atmospheric paleopressure. The existence of >3.7 Gya, ≤$10^2$m-diameter craters should place an upper limit on ancient atmospheric pressure (Popova et al., 2003); we are pursuing this quantitatively.

**Acknowledgements:** We thank Mike Lamb, Ken Farley, Lauren Edgar, Katie Stack, Mathieu Lapôtre, Roman DiBiase, Kirsten Siebach, Ken Farley, Joel Schiengross, Devon Burr, Alexandra Lefort, Robert Jacobsen, and especially Mark Allen, Woody Fischer, Kevin Lewis, and Nick Warner, for discussions. We additionally thank Kevin Lewis and Oded Aharonson for sharing their preprint on cyclic bedding (which was referred to extensively in writing this paper), Oded Aharonson for comments that significantly improved the manuscript, and Mark Allen for reminding us of the importance of resurfacing rate to organic-matter preservation. We thank the HiRISE team for maintaining a responsive public target request program, HiWish, which was useful for this work. DTMs produced for this work, ESP_17548_1740/ESP_019104_1740 (@1m/pixel) and PSP_007474_1745/ ESP_024497_1745 (@2.5m/pixel) are available for unrestricted further use from the corresponding author.

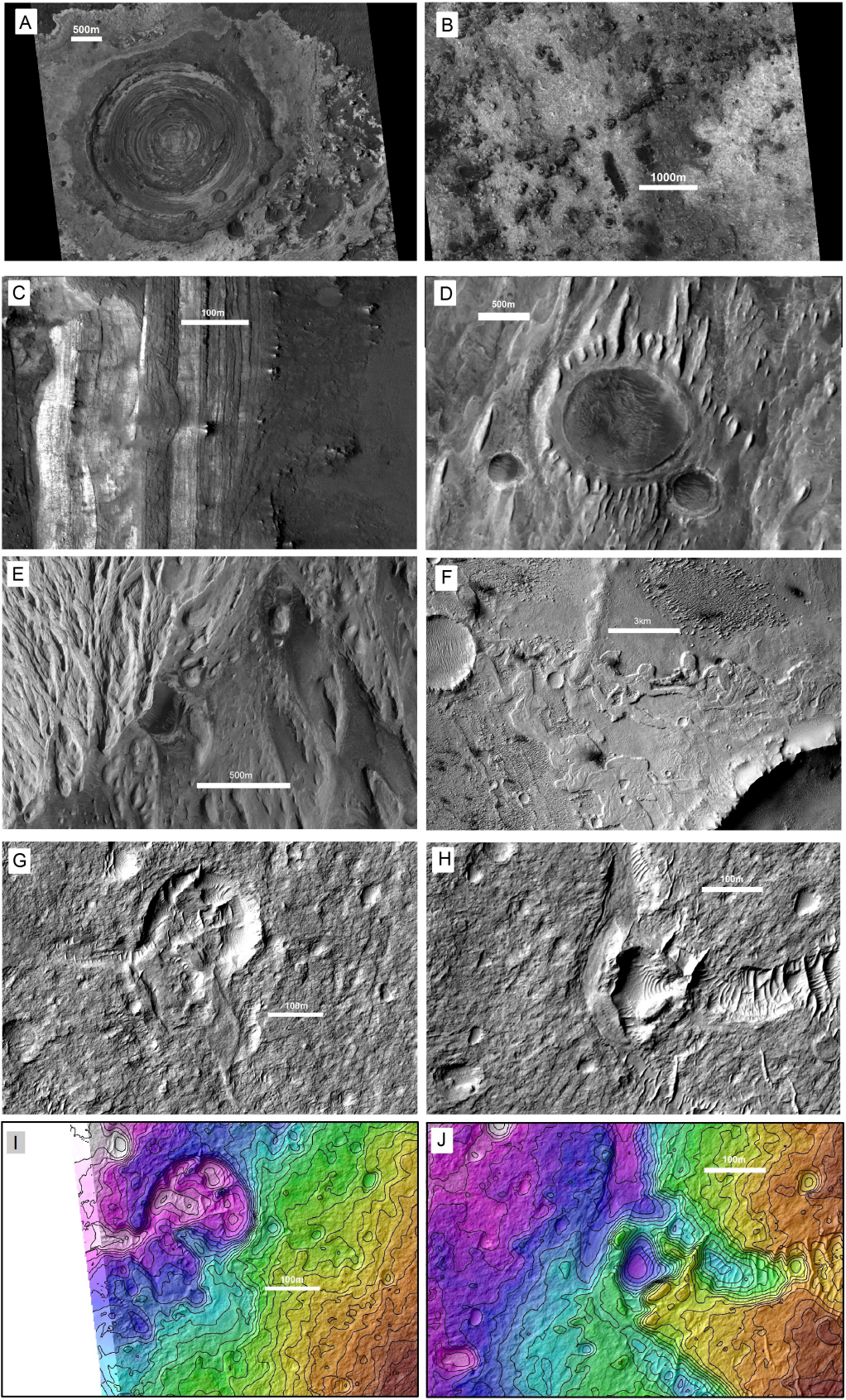

335

**Figure 1.** Varying styles of small-crater preservation on Mars. a) Concentric sediment fill of ~2km-diameter crater, Meridiani. PSP_001981_1825. See Edgett, (2005). b) Secondary-crater chain preserved as rimmed mesas, near Oyama Crater. Rimmed mesas are ~150m diameter. PSP_010882_2040. c) Buried impact exposed in cross-section by sidewall of later crater. Crater is ~200m across. ESP_019664_2035 d) Preservation in inverted relief of crater within Gale Crater's mound (Aeolis Mons). Crater is ~1.2 km across. G05_020054_1749_XN_05S222W. See Thomson et al. (2011). e) Crater being exhumed from beneath the lower unit/upper unit unconformity within Gale Crater's mound (Aeolis Mons). ESP_019988_1750. See Thomson et al. (2011). f) Craters with fresh-appearing ejecta being exhumed from beneath meander belts, Aeolis Dorsa, P15_006973_1742_XI_05S205W; Burr et al., 2010. g) Crater being exhumed from beneath fluvial channel deposit, Aeolis Dorsa. #6 in Supplementary Table, 238 m diameter. ESP_019104_1740. h) Crater draped by fluvial channel deposit, Aeolis Dorsa, ESP_019104_1740. #3 in Supplementary Table, 141 m diameter. i) Crater from (g), but with 1m elevation contours from 1m DTM. DTM is composed of ESP_019104_1740 and ESP_017548_1740. j) Crater from (h), with 1m contours from same DTM.

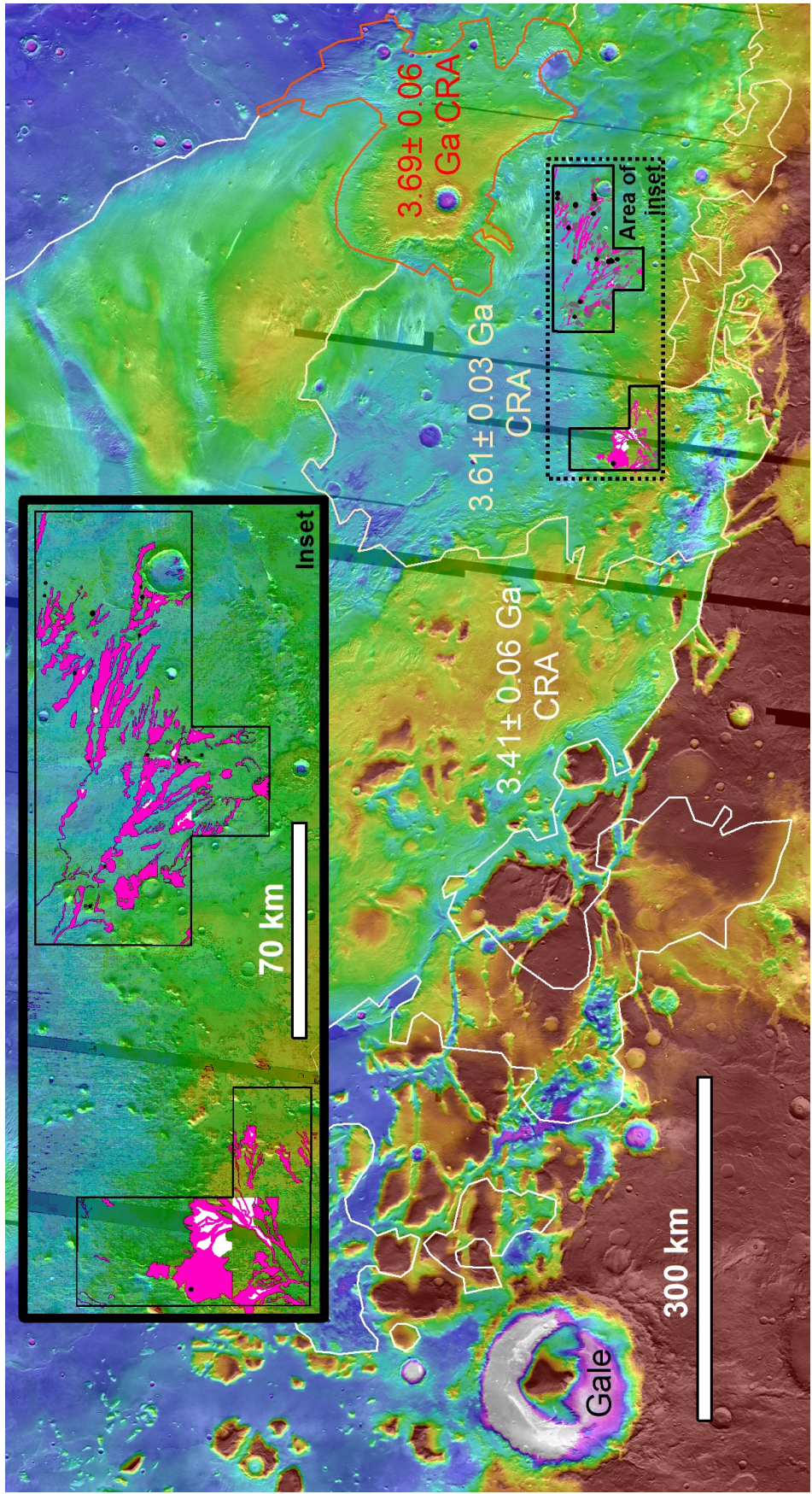

**Figure 2**. Location of study area (black polygons outlining bright pink areas) in relation to Gale Crater and to areas with Crater-Retention Ages (CRAs) determined by Zimbelman & Scheidt (2012). Color scale corresponds to MOLA elevation, is linear, and saturates at -1000m (red) and -3500m (white). Red-outlined area is typically 500m higher than the fluvial-channel deposits. Gale's lower mound accumulated at the same time, within error (Thomson et al., 2011). Inset shows zoom on count region. Within the count region, bright pink areas correspond to counted fluvial-channel deposits, black circles to embedded craters associated with those fluvial deposits, dark gray circles to possible embedded craters, and white areas to "holes" (areas not counted due to poor preservation) within fluvial-channel deposits. Circles representing craters are not to scale. Only a small subset of fluvial-channel deposits was used for this count (Burr et al., 2009).

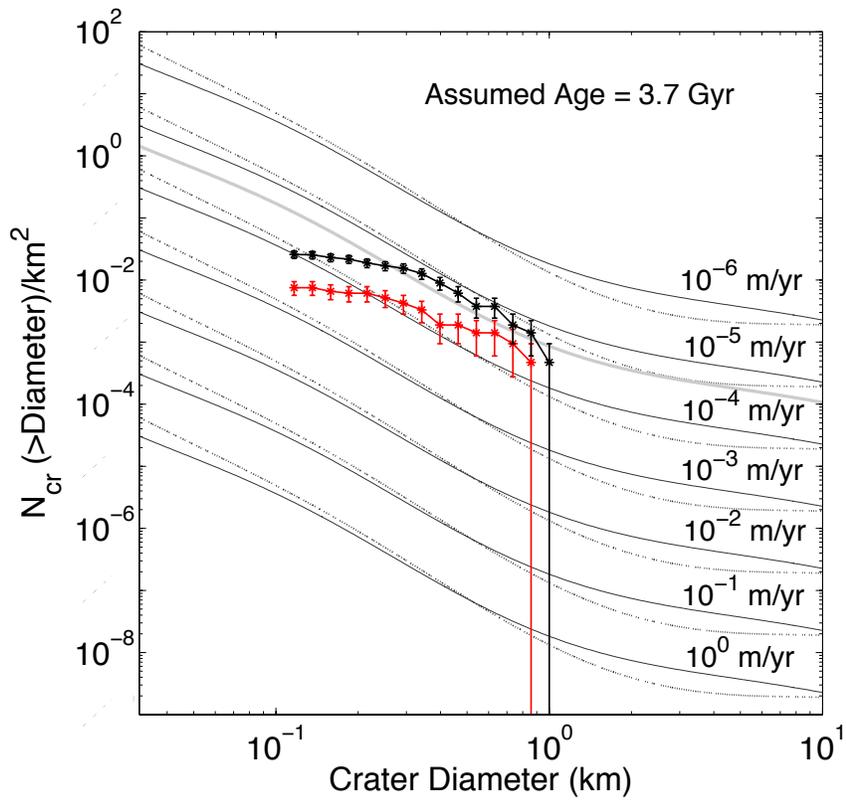

364 **a)**

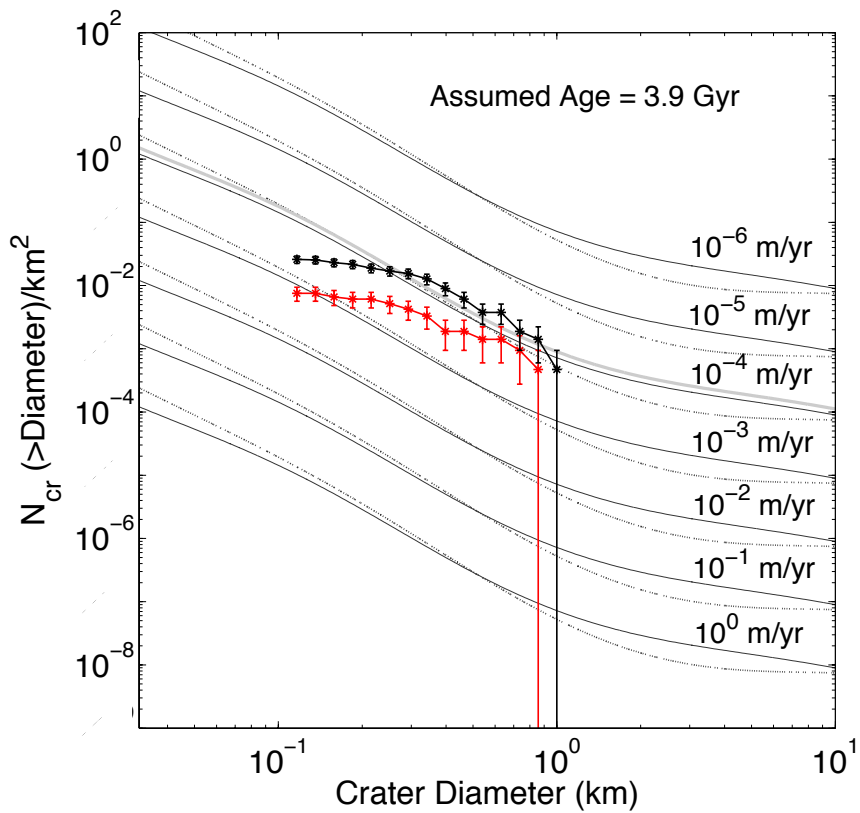

365 **b)**
366

**Figure 3.** Aeolis Dorsa embedded-crater frequencies, plotted against sedimentation-rate curves for (a) assumed age 3.7 Ga and (b) assumed age 3.9 Ga. Red lines with symbols correspond to counts of definite river-crater interactions (Supplementary Table); Black lines with symbols correspond to counts including all candidates. Thin lines are lines of equal sedimentation rate using crater production functions and chronology functions given by Werner & Tanaka, 2011. Thin dotted lines employ "Ivanov" crater production function and "Hartmann & Neukum" chronology function. Thin solid lines employ "Hartmann" crater production function and "Hartmann 2005" chronology function. All function parameters are taken from Werner & Tanaka (2011). Thick gray line corresponds to best-fitting accumulation rate.